\documentclass{article}

\usepackage{axodraw}
\def\nn{\nonumber}
\def\ket#1{| #1 \ra }
\def\bra#1{\la #1}

\def\la{\langle}
\def\ra{\rangle}

\def\l{\left}
\def\r{\right}
\def\nn{\nonumber}

\def\beq{\begin{equation}}
\def\eeq{\end{equation}}
\def\bea{\begin{eqnarray}}
\def\eea{\end{eqnarray}}
\def\barr{\begin{array}}
\def\earr{\end{array}}

\def\ln#1{\log{\l( #1 \r)}}

\begin{document}

\begin{titlepage}
\begin{flushright}
Roma-1347/02\\
SHEP 02-28\\
SISSA 86/02 EP
\end{flushright}
\vskip 0.5cm
\begin{center}
{\Large \bf Finite-Volume Two-Pion Amplitudes \\ \vskip0.2cm in
the \boldmath{$I=0$} Channel} \vskip1cm {\large\bf
C.-J.D.~Lin$^a$, G.~Martinelli$^b$, E.~Pallante$^c$,
C.T.~Sachrajda$^{a}$, G.~Villadoro$^b$}\\ \vspace{.5cm}
{\normalsize {\sl $^a$ Dept. of Physics and Astronomy, Univ. of
Southampton,\\ Southampton, SO17 1BJ, UK. \\ \vspace{.2cm} $^b$
Dip. di Fisica, Univ. di Roma ``La Sapienza" and INFN,\\ Sezione
di Roma, P.le A. Moro 2, I-00185 Rome, Italy.}}\\ \vspace{.2cm}
$^c$ {\sl SISSA, Via Beirut 2-4, 34013, Trieste, Italy.}

\vskip1.0cm {\large\bf Abstract:\\[10pt]} \parbox[t]{\textwidth}{{
We perform a calculation in one-loop chiral perturbation theory of
the two-pion matrix elements and correlation functions of an $I=0$
scalar operator, in finite and infinite volumes for both full and
quenched QCD. We show that major difficulties arise in the
quenched theory due to the lack of unitarity. Similar problems are
expected for quenched lattice calculations of $K \to \pi \pi$
amplitudes with $\Delta I=1/2$. Our results raise the important
question of whether it is consistent to study $K\to\pi\pi$
amplitudes beyond leading order in chiral perturbation theory in
quenched or partially quenched QCD.}}
\end{center}
\end{titlepage}

\section{Introduction}

A precise quantitative evaluation of weak non-leptonic amplitudes
in kaon decays is an enormous challenge for lattice QCD. Although
it has been demonstrated that such a calculation is possible in
principle, a number of major practical difficulties must first be
overcome. These difficulties are related to the construction of
finite matrix elements of renormalized operators from the lattice
bare ones and to the extraction of physical amplitudes, including
final state interaction phases, from Euclidean correlation
functions. For the latter problem, it has been demonstrated that
it would be possible in principle to obtain the physical
amplitudes by performing unquenched simulations with physical
quark masses on lattice volumes large enough  to have
discretization errors and  finite size effects under
control~\cite{LL,noi}.  At present however, it is not possible to
perform unquenched simulations on such large volumes and therefore
a certain number of approximations are necessary. One of the main
approximations (in addition to quenching) consists in working with
unphysical quark masses and/or external meson momenta, and
estimating the physical amplitudes by extrapolating to the
physical point. A key element of our strategy in evaluating
$K\to\pi\pi$ matrix elements is the use of Chiral Perturbation
Theory ($\chi$PT) at next-to-leading order (NLO)~\cite{spqr}. In a
recent paper~\cite{spqrdi32logs}, we have presented the relevant
formulae for $\Delta I=3/2$ transitions on finite and infinite
volumes, in the full theory and in the quenched approximation. Our
results show explicitly that all corrections which vanish as
inverse powers of the volume can be eliminated by using the
methods introduced in refs.~\cite{LL,noi}~\footnote{The
infinite-volume limit, $L\to\infty$ (where $L$ is the length of
each spatial dimension of the lattice), is to be taken at fixed
physics, i.e. at fixed two-pion energy $W$.}. The remaining finite
volume corrections are exponentially small (of ${\cal O}(e^{-m
L})$). At the NLO in $\chi$PT this is true also in the quenched
approximation.

\par In this paper we present the results obtained at NLO in
$\chi$PT for matrix elements with an $I=0$ two-pion final state.
We study these matrix elements in order to illustrate the main
features present in $\Delta I=1/2$ $K\to\pi\pi$ transitions in
quenched QCD. For these decays, the lack of unitarity of the
quenched theory leads to a number of problems which need to be
solved in order to understand the volume dependence and to extract
the amplitudes. The main consequences of quenching, due to the
lack of unitarity, can be summarized as follows:
\begin{itemize}
\item  the final state interaction phase is not universal, since
it depends on the operator used to create the two-pion state. This
is not surprising, since the basis of Watson's theorem is
unitarity; \item the L\"uscher quantization
condition~\cite{luscher} for the two-pion energy levels in a
finite volume does not hold~\cite{BG95}; \item a related
consequence is that the Lellouch-L\"uscher (LL) relation between
the absolute value of the  physical amplitudes and the finite
volume matrix elements~\cite{LL,noi} is no longer valid. It is
therefore not possible to take the infinite volume limit at
constant physics, namely with a fixed value of $W$; \item whereas
it is normally possible to extract the lattice amplitudes by
constructing suitable time-independent ratios of correlation
functions, this procedure fails in the quenched theory as
explained in section~\ref{subsec:quencheddiscussion}. In
particular the time dependence of correlation functions
corresponding to different operators which create the same
external state is not the same; \item in addition to the usual
exponential dependence on the time intervals, the presence of the
double pole corresponding to the incomplete $\eta^\prime$
propagator generates, at one-loop order in $\chi$PT, terms in the
Euclidean correlation functions which depend linearly,
quadratically or cubically on the time~\cite{Golterman:1999hv}.
Unlike the corrections which shift the two-pion energy in a finite
volume~\cite{spqrdi32logs}, these terms do not exponentiate and
may cause practical problems in the extraction of the finite
volume matrix elements. A related problem is the appearance, at
fixed $L$, of corrections linear or cubic in
$L$~\cite{Golterman:1999hv, BG95}.
\end{itemize}

The dependence on the $\eta^\prime$ parameters can be removed by
working in partially quenched QCD, where the $\eta^\prime$ is
heavy and decouples from the light Goldstone boson sector,
although, in general, residual double poles will remain. All the
problems originating from the lack of unitarity (denoted as the
{\it unitarity problem} in the following) would, however, remain
the same. Unitarity is recovered from partially quenched QCD only
in the limit when the number of sea and valence quarks is equal
and their masses are equal. This corresponds to full QCD.

\par In view of the difficulties listed above, it may be questioned
whether it is possible to obtain $K\to\pi\pi$ decay amplitudes
beyond leading order in $\chi$PT in quenched or partially quenched
QCD. In the absence of a solution to the problems encountered and
discussed in this paper, we would be limited to extracting the
effective couplings (the low-energy constants) corresponding to
the operators in the weak Hamiltonian at lowest order in the
chiral expansion from $K \to \pi$ matrix elements computed in
lattice simulations. The absence of unitarity is intrinsic to
quenched and partially quenched QCD, and we do not have a solution
to the unitarity problem. Nevertheless it is tempting to speculate
whether there might not be a possible pragmatic way to proceed, in
spite of the failure of Watson's theorem. Indeed the key point is
that, in the quenched case, the two-pion state is no longer an
eigenstate of the strong interaction Hamiltonian. The eigenstates
of the Hamiltonian would be, formally, linear combinations of
physical pions and unphysical mesons composed of the
pseudo-fermion fields. The latter, however, have the wrong
spin-statistic properties and for this reason unitarity breaks
down.   We speculate that it might be possible to recover a
variation of Watson's theorem, and of the LL formula in finite
volumes, by a suitable re-interpretation of the quenched theory,
for example by using the replica method of ref.~\cite{ds}, working
in the basis of Hamiltonian eigenstates. However, we stress that
this is only a speculation and we will report on the conclusions
of our investigations of this important question in a future
paper.

\par Since all the major difficulties arising from the {\it unitarity
problem} depend only on the quantum numbers of the operators and
final state, for the sake of illustration we discuss in this paper
matrix elements of the form $\langle \pi\pi  |S| 0\rangle$, where $S$
is a scalar and isoscalar operator which can annihilate the
two-pion state. We also discuss the properties of the correlation
functions from which such matrix elements are obtained. The
discussion can readily be extended to $K\to\pi\pi$ matrix elements
of $\Delta I=1/2$ operators of the effective weak
Hamiltonian~\cite{hw}.  The results for the $\Delta
I=1/2$ amplitudes in one loop $\chi$PT on finite and infinite
volumes, in the full theory and in the quenched approximation,
will be presented in a forthcoming publication~\cite{future}. In
our calculation we have used the formulation of the quenched
chiral Lagrangian  introduced in ref.~\cite{bg}, using the
conventions and notation presented in section 3 of
ref.~\cite{spqrdi32logs}. The scalar operator is defined by
\begin{equation}
S = \mbox{tr} \Bigl[ \Sigma + \Sigma^{\dagger} \Bigr]  , \eeq in
the full theory and as
\begin{equation}
S^q =  \sum_{i=1,3} \Bigl[\Sigma^q + \Sigma^{q\,\dagger}
\Bigr]_{ii} , \eeq in the quenched theory, where the trace (sum)
is taken over the indices of the chiral group $SU(3)_L\otimes
SU(3)_R$ (graded group $SU(3|3)_L\otimes SU(3|3)_R$). In the
quenched case the field $\Sigma^q$ is the graded extension of the
standard field $\Sigma$ of the full theory.

The main results of our calculations are presented in the four
appendices \ref{app:sa}--\ref{app:fvqsa}, which contain the
following:
\begin{enumerate}
\item  the NLO expression for the matrix element of the scalar operator
in full QCD and in infinite volume, $\bra{\pi^+\pi^-}|  S
\ket{0}$. This is given in eq.~(\ref{eq:sft});
\item  the expression for the corresponding quantity in the quenched
theory $\bra{\pi^+\pi^-}| S^{q}\ket{0}$, presented in
eq.~(\ref{eq:sqt});
\item  the NLO result for the correlation function
$ \bra{0}| \pi^+_{- \vec q}(t_1) \pi^-_{\vec q}(t_2) S(0)\ket{0}$
in a finite volume and in full QCD. This is given in
eq.~(\ref{eq:scft});
\item  the corresponding correlation function in the quenched
theory, $ \bra{0}| \pi^+_{- \vec q}(t_1) \pi^-_{\vec q}(t_2)
S^q(0)\ket{0}$, presented in eq.~(\ref{eq:scqt}).
\end{enumerate}
In the above $S(0) \equiv S(\vec x=0,t=0)$ with the corresponding
definition of $S^q(0)$. The expressions for the matrix elements
are given in Minkowski space, whereas those for the correlation
functions are presented in Euclidean space. In the correlation
functions we have used the following definition for the Fourier
transform of the fields
\begin{equation}
\pi_{\vec q}(t)  = \int \; d^{\hspace{0.8pt}3}\hspace{-0.8pt}x\
\pi(\vec x , t)\, e^{i \vec q \cdot \vec x} \, . \eeq The
correlation functions depend on the choice of $t_1$ and $t_2$. In
this paper, for purposes of illustration, we present the results
for the two cases $t_1=t_2$ and $t_1\gg t_2$.
\par In section \ref{sec:discussion} we discuss
the physical interpretation  of the expressions obtained at
one-loop in $\chi$PT for the different cases and the implications
for the relation between finite volume Euclidean correlation
functions and physical amplitudes (including final state
interaction phases)~\cite{LL,noi}. The relevant expressions and
the technical details can be found in the appendices.

\section{Discussion of the One-loop Calculations}
\label{sec:discussion}

In this section, we discuss the extraction of the physical
amplitudes from finite-volume Euclidean correlation functions,
using the results obtained in one-loop $\chi$PT. We first consider
the unquenched case, where we explicitly check the validity of the
LL relation~\cite{LL,noi}, derived using general properties of
quantum mechanics and field theory.

\subsection{Extraction of the Physical Amplitude from the Scalar
Correlation Function in Full QCD}

We begin our discussion from the correlation function of the
scalar operator with two pion fields, in the full theory  at
finite volume, $ \bra{0}| \pi^+_{- \vec q}(t_1) \pi^-_{\vec
q}(t_2) S(0)\ket{0}$. The tree-level and one-loop diagrams are
shown in figs. 1t and 1z, 1a and 1b. Final state interactions, and
consequently power-like finite volume corrections, are only  given
by the diagram in fig.~1b~\footnote{The evaluation of this diagram
for $I=2$ final states was explained in some detail in section 4
of ref.~\cite{spqrdi32logs}. We therefore do not present a
description of the calculation for $I=0$ final states, but limit
the discussion to the implications of the results.}. The NLO
expression for the correlation function is given in
eq.~(\ref{eq:scft}). The corresponding Minkowski amplitude in
infinite volume is given in eq.~(\ref{eq:sft}).
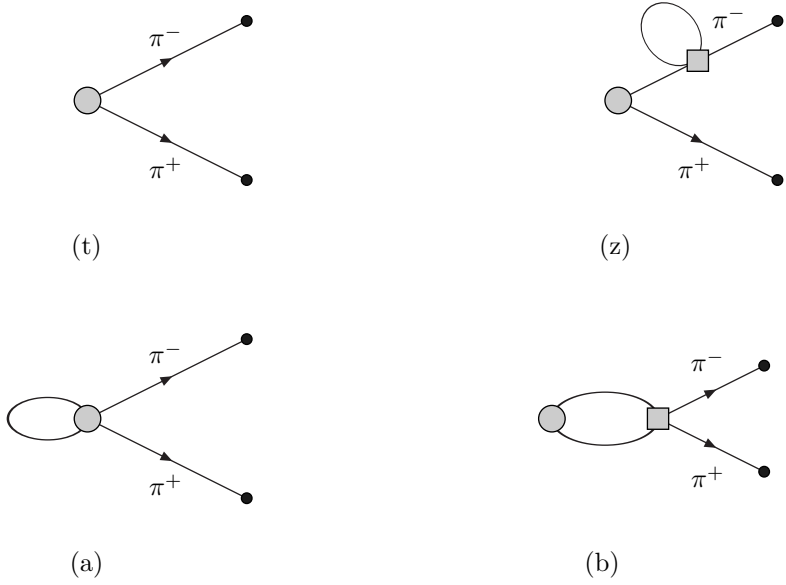
\begin{figure}
\begin{center}
\begin{picture}(400,250)(0,-65)
\ArrowLine(75,30)(135,60) \ArrowLine(75,30)(135,0)
\Oval(60,30)(8,15)(0) \GCirc(75,30){5}{0.8} \GCirc(135,60){2}{0.1}
\GCirc(135,0){2}{0.1}
\Text(105,51)[b]{$\pi^-$}
\Text(105,9)[t]{$\pi^+$} \Text(75,-20)[t]{(a)}
\ArrowLine(290,30)(330,50)
\ArrowLine(290,30)(330,10) \Oval(270,30)(10,20)(0)
\GCirc(330,50){2}{0.1}
\GCirc(330,10){2}{0.1}
\Text(310,48)[b]{$\pi^-$}\Text(310,12)[t]{$\pi^+$}
\GCirc(250,30){5}{0.8} \GBoxc(290,30)(8,8){0.8}
\Text(270,-20)[t]{(b)}
\ArrowLine(75,150)(135,180)
\ArrowLine(75,150)(135,120)
\GCirc(75,150){5}{0.8}
\GCirc(135,180){2}{0.1}
\GCirc(135,120){2}{0.1}
\Text(105,171)[b]{$\pi^-$}
\Text(105,129)[t]{$\pi^+$} \Text(75,100)[t]{(t)}
\ArrowLine(275,150)(335,180)
\ArrowLine(275,150)(335,120)
\Oval(295,175)(13,10)(40)
\GCirc(275,150){5}{0.8}
\GCirc(335,180){2}{0.1} \GBoxc(305,165)(8,8){0.8}
\GCirc(335,120){2}{0.1}
\Text(318,178)[b]{$\pi^-$}
\Text(305,129)[t]{$\pi^+$} \Text(275,100)[t]{(z)}
\end{picture}
\caption{\it Tree level (t) and one-loop $\chi$PT diagrams for the
$\bra\pi^-\pi^+|S\ket0$ amplitude and $\bra0 | \pi^-\pi^+S \ket0$
correlation function. The grey circle represents the scalar source
while the squares are strong vertices. \label{graphs}}
\end{center}
\end{figure}
In this case we denote the contribution of the  diagrams in
fig.~1z,~1a and 1b by $I_z$, $I_a$ and $I_b$ respectively and
define the relative one-loop correction to the infinite volume
amplitude as ${\cal A}_\infty= 1/(4\pi f)^2\,(I_z + I_a +I_b)$,
see eqs.~(\ref{eq:sft}) and (\ref{eq:spiega}). Final state
interactions are encoded in the function $A(m)$ introduced in
eq.~(\ref{eq:adm}).

In a finite volume the corrections to the amplitude from the
diagrams in figs. 1z and 1a ($I_z$ and $I_a$ respectively) are the
same as in infinite volume up to  exponentially small terms (in
the volume) that will be neglected in the following. The diagram
in fig. 1b gives a correction, $I_b(t_1,t_2)$, which is a function
of the time coordinates of the interpolating operators which
annihilate the two pions (the two-pion sink).

At lowest order (obtained by setting
$I_{z}=I_{a}=I_{b}(t_1,t_2)=0$), the time dependent factor
\begin{equation} \frac{e^{-Et_1}}{2E}\frac{e^{-Et_2}}{2E}  \, ,
\label{eq:time} \eeq can be removed by dividing by the two-pion
propagator in the free theory, and in this way the required matrix
element can be obtained. In equation~(\ref{eq:time}) $E$ is the
energy of each of the pions, $E=\sqrt{\vec q^{\hspace{2pt}2} +
m_{\pi}^{2}}$.

At one-loop order the relevant finite-volume correction to the
correlation function of eq.(\ref{eq:scft}) is therefore given by
$I_{b}(t_{1},t_{2})$ which is presented explicitly in
eq.~(\ref{eq:ibu}). We rewrite $I_{b}(t_{1},t_{2})$ as
$I_{b}(t_{1},t_{2})= {\rm Re}(I_{b}) +{\cal T}(t_{2}) + {\cal
R}(t_1,t_2)$, where $I_b$ is the corresponding infinite-volume
one-loop contribution to the matrix element (see
eq.~(\ref{eq:spiega})\,) and we now discuss the significance of
the terms ${\cal T}$ and ${\cal R}$.

${\cal T}(t_{2})$ contains the one-loop corrections which are
multiplied by the \textit{correct} time dependence,\\
$\exp(-Wt_2)\exp(-E(t_1-t_2))$ (after exponentiation), where
$W=2E+\Delta W$ and $\Delta W$ is the shift of the two-pion energy
due to interactions in the finite volume~\cite{spqrdi32logs}. We
can readily extract $\Delta W$ from the coefficient of $t_2$ in
the expression for ${\cal T}(t_{2})$ in eq.~(\ref{eq:sopra}),
\begin{equation}\label{eq:deltaw}
\Delta W = -\frac{\nu}{4f^2L^3}(8-\frac{m_\pi^2}{E^2}) \, . \eeq
It is then straightforward to check that eq.~(\ref{eq:deltaw})
reproduces the well known result for the scattering length
$a_{0}^{I=0}$~\cite{gl}. When  the two pions are at rest
eq.~(\ref{eq:deltaw}) gives
\begin{equation}
W=  2m_\pi - \frac{4 \pi a_{0}^{I=0}}{m_{\pi}  L^{3}} \, , \eeq
 where  $a_{0}^{I=0} = 7 m_{\pi} /(16 \pi
f^{2}_{\pi})$.

${\cal T}(t_{2})$ also contains the finite volume corrections to:
\begin{enumerate}\vspace{-10pt}
\item the matrix element of the scalar operator. These are given
by the one-loop component of the LL-factor relating the
infinite-volume and finite-volume amplitudes, ${\cal M}_\infty$
and ${\cal M}_V$ respectively, ($ |{\cal M}_{\infty}|^2 = LL
\times |{\cal M}_{V}|^2$);
\item the two-pion sink used to annihilate the pions created by
the scalar source. We refer to these as \textit{Forward Time
Contributions} or $FTCs$. They are presented explicitly in
eq.~(\ref{eq:varie}).
\end{enumerate}
The $FTCs$ are eliminated by dividing $\bra{0}| \pi^+_{- \vec
q}(t_1) \pi^-_{\vec q}(t_2) S(0)\ket{0}$ by (the square root of) a
suitable $\pi\pi$ correlation function~\cite{spqrdi32logs}
\begin{eqnarray}\label{eq:4pi}
G_{\pi\pi \to \pi\pi}(t_{1},t_{2}) &\equiv&  \sum_{ \vec{p},\vec q
;\ |\vec p\,|=|\vec q\,|,\ |\vec q\,| \textrm{\ fixed} } \la 0|\,\pi_{\vec
q\,}(t_1)\pi_{-\vec
q\,}(t_2)\,\pi_{-\vec{p}\,}^{\dag}(-t_2)\pi_{\vec{p}\,}^{\dag}(-t_1)\,|0\ra
\\ \rule[-5pt]{0pt}{30pt} &=& \nu \, \frac{e^{-2 Wt_2}}{(2E)^2}L^6
\l(1+\l(1-\frac{3m_\pi^2}{8E^2}\r) \frac{2\nu}{3f^2EL^3}\r) \hfill
\hspace{1.07in}\textrm{for }t_1=t_2 \nn \\ \rule[-5pt]{0pt}{30pt}
&=&\nu \, \frac{e^{-2 Wt_2-2E(t_1-t_2)}}{(2E)^2}L^6
\l(1+\l(1-\frac{3m_\pi^2}{8E^2}\r)
\frac{\nu}{3f^2EL^3}\r)\hspace{0.5in}\textrm{for }t_1 \gg
t_2\,.\nn \eea Note that in the unquenched case the finite-volume
energy $W$ appearing in eq.~(\ref{eq:4pi}) is, as expected, the
same as in \beq
 \bra{0}| \pi^+_{- \vec q}(t_1) \pi^-_{\vec q}(t_2) S(0)\ket{0}
\propto
 e^{-E(t_1-t_2)} e^{-W t_2}  \, , \eeq
because we are considering the same final state. As shown below,
this is not true  in the quenched and in the partially quenched
theories.

Finally ${\cal R}(t_1,t_2)$ corresponds to contributions to the
correlation function whose time dependence is governed by energies
different from $W$ and which therefore can be eliminated by
studying this time dependence~\cite{spqrdi32logs,ciuchini}. Since
it is not relevant to our discussion, ${\cal R}(t_1,t_2)$ will be
neglected in the formulae given in the appendices.

\subsection{The Scalar Correlation Function in the Quenched
Theory}\label{subsec:quencheddiscussion}

In one-loop $\chi$PT with $I=2$ final states (as for example in
$\Delta I=3/2$ $K \to \pi\pi$ transitions) we can follow the path
outlined above for full QCD also in the quenched theory. This is
not the case however, when we consider the infinite-volume
amplitude or the finite-volume correlation function of the
isosinglet scalar operator with two pion fields in the quenched
theory. The one-loop expression for the amplitude $
\bra{\,\pi^+\pi^-\,}|\, S^q\,\ket{\,0\,}$ is given in
eq.~(\ref{eq:sqt}) and for the correlation function $
\bra{\,0\,}|\,\pi^+_{-\vec q\,}(t_1) \pi^-_{\vec q}(t_2)
S^q(0)\,\ket{\,0\,}$ in eq.~(\ref{eq:scqt}). In this case, even in
infinite volume there are severe difficulties, for example the
imaginary part of the amplitude given in eq.~(\ref{eq:sqt}),
corresponding to the final state interactions, depends on the
scalar operator and diverges at threshold, i.e. when $s \to 4
m_\pi^2$~\cite{BGLSU,Colangelo:1997ch}. We now consider what
happens for the correlation function in a finite volume.

\par For illustration, let us start by neglecting the contribution of the
$\eta^\prime$ double pole by setting $m_{0}=0$ and $\alpha=0$ (they are
defined in eq.~(\ref{eq:lstrongq})\,). For
simplicity we also neglect the terms proportional to $v_{1,2}$
($v_{1,2}$ are also defined in eq.~(\ref{eq:lstrongq})\,). This is similar
to the situation encountered in the partially quenched case where
the $\eta^\prime$ is heavy, but unitarity is violated. In this
case all the $a^{(i)}_{11}$ defined in Appendix~\ref{app:fvqsa} satisfy
$a^{(i)}_{11}=0$ and for $E < m_K < m_{\bar s s}$
only the first line of ${\cal T}^q(t_2)$ in eq.~(\ref{eq:calTq})
contributes \bea {\cal T}^q(t_2)= -\frac{E^{2}}{2 f^{2}} \l[
-\frac{2\nu t_2}{E^2
L^3}+\l(-\frac{\nu}{3E^3L^3};-\frac{\nu}{6E^3L^3}\r)
    +\l(\frac{z(0)}{2E^3L^3}-\frac{z(1)}{2\pi^2EL}\r)\r] \label{eq:tq} \, .\eea
The terms $\l(...;...\r)$ are the forward-time contributions
($FTCs$) respectively for the two cases $t_1 \gg t_2$ and $t_2=t_1$.

\par The first term in eq.~(\ref{eq:tq}) is the shift in the energy of the
two-pion final state. This shift,
\begin{equation} \Delta W =  -
\frac{\nu }{f^2 L^3} \end{equation} is different from that
obtained from the $\pi \pi \to \pi\pi$ correlation function in
eq.~(\ref{eq:4pi}), which is
\begin{equation} \Delta W =  -
\frac{\nu }{4 f^2 L^3}\l(8 -\frac{m^{2}_{\pi}}{E^{2}}\r)  \nn  \,
.\end{equation} The latter in fact is the same in the quenched and
unquenched theories, since, at this order, it is given by a tree
diagram. Once the contribution from the $\eta^\prime$ double pole
is included, at this order there are quadratic and cubic terms in
$t_2$ present in ${\cal T}^q(t_2)$, but not in the
$\pi\pi\to\pi\pi$ correlation function.

The second term in eq.~(\ref{eq:tq}) should be cancelled when
extracting the matrix element of the scalar operator by dividing
by the square root of the $\pi\pi\to\pi\pi$ correlation function:
\begin{equation}\label{eq:ratio} |\bra{\pi\pi}|S^{q} \ket{0}_{V}| =
\frac{ \bra{0}| \pi^+_{- \vec q}(t_1) \pi^-_{\vec q}(t_2) S^{q}(0)
\ket{0} } {\sqrt{G_{\pi\pi \to \pi\pi}(t_{1},t_{2})} } \, ,  \eeq
since it is the finite volume correction to the sink operator used
to annihilate the two pions. One  can readily verify that the
cancellation does not occur, unlike in the unquenched theory. Thus
the power corrections in $1/L$ to $|\bra{\pi\pi}| S^{q}
\ket{0}_{V}|$ are not those expected on the basis of the
Lellouch-L\"uscher formula.
\section*{Conclusion}
In this paper we have shown that the standard strategy for
extracting the amplitude and the relative phase from finite-volume
calculations of correlation functions fails in the quenched theory
for $I=0$ two-pion final states. We encounter the same problems
when computing the matrix elements of the operators relevant for
$\Delta I=1/2$ $K\to\pi\pi$ transitions at one-loop order in
$\chi$PT~\cite{future}.

At present we do not know whether it might be possible to recover
a modification of Watson's theorem and of the LL formula in finite
volumes, which would allow for a consistent determination of $I=0$
two-pion matrix elements beyond leading order in quenched
$\chi$PT. We are currently investigating the possibility of
performing a suitable analytic continuation in the numbers of
flavours, using the replica method of ref.~\cite{ds} at one-loop
order in $\chi$PT.

\section*{Acknowledgements}
We thank Massimo Testa for helpful discussions and Maarten
Golterman for comments on the manuscript.

This work was supported by European Union grant
HTRN-CT-2000-00145. CJDL and CTS acknowledge support from PPARC
through grants PPA/G/S/1998/00529 and PPA/G/O/2000/00464. EP was
supported in part by the Italian MURST under the program
\textit{Fenomenologia delle Interazioni Fondamentali}.

\begin{center}
{\LARGE\textbf{Appendices}}
\end{center}
\appendix
\section{The Scalar Amplitude in Full QCD}
\label{app:sa} To fix the conventions we start by writing the
chiral Lagrangian used in our calculations in the full QCD:
\begin{equation}
{\cal L}_{\textrm{\scriptsize{strong}}}=\frac{f^2}{8}{\textrm
tr}\left[(\partial_\mu \Sigma^\dagger)\,(\partial^\mu\Sigma) +
\Sigma^\dagger\chi + \chi^\dagger\Sigma\,\right]\,,
\label{eq:lstrong}\end{equation} where the decay constant $f$ is
normalized in such way that $f_\pi \sim 132$~MeV.
\par  The one-loop matrix element of the scalar operator
in full QCD in infinite volume is given by
\begin{equation}
\bra{\pi^-(\vec q\,)\,\pi^+(-\vec q\,)}\,|\, S\, \ket{\,0\,}\equiv
-\frac{8}{f^2}\l[1+{\cal A}_{\infty}\r] =
-\frac{8}{f^2}\l[1+\frac{1}{(4\pi f)^2}\l(I_z+I_a+I_b\r)\r] \, ,
\label{eq:sft} \eeq where
\begin{eqnarray}
I_z&=& -\frac{2}{3}m_K^2\ln{\frac{m_\pi^2}{m_K^2}}+\frac{2}{3}
\l(m_K^2+2m_\pi^2\r)\ln{\frac{m_\pi^2}{\mu^2}} \, , \nn \\
I_a&=&-\frac{m_\eta^2}{3}\ln{\frac{m_\eta^2}{\mu^2}}-\frac{5m_\pi^2}{3}
\ln{\frac{m_\pi^2}{\mu^2}}-\frac{4m_K^2}{3}\ln{\frac{m_K^2}{\mu^2}}\,,\nn\\
I_b&=&(m_\pi^2-2s)A(m_\pi)-s A(m_K)-\frac{m_\pi^2}{3}A(m_\eta) \nn
\\ & &+\l(\frac{7}{3}m_\pi^2-2s\r)\ln{\frac{m_\pi^2}{\mu^2}}
    +\l(\frac{2}{3}m_K^2-s\r)\ln{\frac{m_K^2}{\mu^2}} \nn \\
& &
-\frac{m_\pi^2}{3}\ln{\frac{m_\eta^2}{\mu^2}}+\l(3s-\frac{2}{3}m_\pi^2\r)
\,.\label{eq:spiega} \eea In the above expressions $\mu$ is the
renormalization scale, $s = (p_{\pi^-} +p_{\pi^+})^{2}$ is the
square of the two-pion center of mass energy and
\begin{equation} A(m)\equiv
\sqrt{1-4\frac{m^2}{s}}\l(\ln{\frac{1+\sqrt{1-4\frac{m^2}{s}}}{1-\sqrt{1-4\frac{m^2}{s}}}}-i\pi
\theta\l(1-4\frac{m^2}{s}\r)\r)\, . \label{eq:adm} \eeq

To the one-loop corrections listed in the above equations, we
should add those proportional to the Gasser-Leutwyler
coefficients, $L_{i}$~\cite{gl}, appearing at NLO in the strong
interaction chiral Lagrangian. These effects, combined with
$I_{z}$, are reabsorbed into the renormalization of the decay
constant, leading to the replacement of the factor $1/f^{2}$ by
$1/f_{\pi}^{2}$ in the matrix elements of the scalar operator.
Since these terms do not affect the finite volume corrections
which are the object of the present study, we will not discuss
them further here. Similarly the matrix elements of the
(higher-dimensional) counter-terms of the scalar density do not
contribute to the finite volume effects and will not be
considered.

The s-wave phase shift for the $I=0$ two-pion state at one-loop in
$\chi$PT is given by \beq \delta(s)=\frac{2s-m_\pi^2}{16\pi
f^2}\sqrt{1-4\frac{m_\pi^2}{s}}\, . \eeq For $s \ge 4 m_\pi^2$, we
have ${\rm Arg}\l[\bra{\pi^-(\vec q\,)\pi^+(-\vec q\,)}| S
\ket{0}\r] = \delta(s)$. Note that the amplitude remains finite at
threshold, i.e. as $s \to 4 m_\pi^2$.

\section{The Scalar Amplitude in Quenched QCD}
\label{app:qsa}

In this appendix we discuss the matrix element of the scalar
density in quenched QCD. The quenched chiral Lagrangian used in
our calculations is:
\begin{eqnarray}
{\cal
L}_{\textrm{\scriptsize{strong}}}^{q}&=&\left(\frac{f^2}{8}+\frac{v_1}{2}\Phi_0^2
\right)\textrm{str}\,\left[(\partial_\mu\Sigma^{q\,\dagger})\,
(\partial^\mu\Sigma^q)\right] +
\left(\frac{f^2}{8}+\frac{v_2}{2}\Phi_0^2
\right)\textrm{str}\,\left[\Sigma^{q\,\dagger}\chi +
\chi^\dagger\Sigma^q\,\right]\nn\\ &&\hspace{1.5in} -
m_0^2\Phi_0^2+\alpha\,(\partial_\mu\Phi_0)(\partial^\mu\Phi_0)\,,
\label{eq:lstrongq}\end{eqnarray} with the super-$\eta^\prime$ field, $\Phi_0=f/2i$
str$[\log{\Sigma^q}]/\sqrt{6}$.
\par The one-loop matrix element of the scalar operator
in the quenched theory, at infinite volume, is given by
\beq
\bra{\pi^+\pi^-}| S^{q} \ket{0}\equiv -\frac{8}{f^2}\l[1+{\cal
A}^q_{\infty} \r] =-\frac{8}{f^2}\l[1+\frac{1}{(4\pi
f)^2}\l(I^q_z+ I^q_a+I^q_b\r)\r] \, , \label{eq:sqt}
\eeq
where
\bea
I_z^q&=&0,\nonumber\\
I_a^q&=&-\frac{2m_\pi^2}{3}\ln{\frac{m_\pi^2}{\mu^2}}-\frac{m_K^2}{3}\ln{\frac{m_K^2}{\mu^2}}
    -\frac{2 m_0^2}{3}\l[1+\ln{\frac{m_\pi^2}{\mu^2}}\r]+\frac23\alpha\, m_\pi^2\l
    [1+2\ln{\frac{m_\pi^2}{\mu^2}}\r]\,, \nn \\
I_b^q&=&s\l(\frac{3}{2}-A(m_\pi)-\frac{1}{2}A(m_K)\r)+\l(\frac{2}{3}m_\pi^2-s\r)\ln{\frac{m_\pi^2}{\mu^2}}
    +\l(\frac{1}{3}m_K^2-\frac{s}{2}\r)\ln{\frac{m_K^2}{\mu^2}} \nn \\
& &+\frac{2 m_\pi^2\l( v_2-v_1 \r)+s v_1}{3}\l(A(m_{{\bar
s}s})+2A(m_\pi)+ 2\ln{\frac{m_\pi^2}{\mu^2}}+\ln{\frac{m_{{\bar
s}s}^2}{\mu^2}}-3\r)\nn\\ &
&+m_0^2\l[\frac{4m_\pi^2}{3(s-4m_\pi^2)}A(m_\pi)\r]
    +\alpha\frac{4m_\pi^2}{3}\l[\ln{\frac{m_\pi^2}{\mu^2}}-1+\frac{s-5m_\pi^2}{s-4m_\pi^2}A(m_\pi)\r]\nn \\
& &+m_0^4\l[\frac{m_\pi^2}{18(m_K^2-m_\pi^2)^2}\l(A_{10}-A(m_{{\bar s}s})\r)-\frac{8m_\pi^2}{9s(s-4m_\pi^2)}+
      \frac{m_\pi^2}{9s(m_K^2-m_\pi^2)}\ln{\frac{m_{{\bar s}s}^2}{m_\pi^2}}\r.\nn\\
& &\
-\l.\frac{m_\pi^2[32m_\pi^6+16m_K^4(2m_\pi^2-s)-8m_\pi^2s^2+s^3-32m_K^2
m_\pi^2(2m_\pi^2-s)]}{18s(m_K^2-m_\pi^2)^2(s-4m_\pi^2)^2}A(m_\pi)\r]\nn\\
& &+\alpha m_0^2
\l[\frac{16m_\pi^4}{9s(s-4m_\pi^2)}+\frac{m_\pi^2}{9(m_K^2-m_\pi^2)^2}\Bigl(m_{{\bar
s}s}^2 A(m_{{\bar s}s})-m_K^2A_{10}\Bigr)
      -\frac{m_\pi^2(2m_K^2-s)}{9s(m_K^2-m_\pi^2)}\ln{\frac{m_{{\bar s}s}^2}{m_\pi^2}} \r.\nn \\
& &\
\l.+\frac{m_\pi^2[8(m_K^2-m_\pi^2)^2(4m_\pi^4+2m_\pi^2s-s^2)+m_\pi^2s(s-4m_\pi^2)^2]}
{9s(m_K^2-m_\pi^2)^2(s-4m_\pi^2)^2}A(m_\pi)\r] \nn \\ &
&+\frac{2m_\pi^2}{3}\alpha^2\l[1-\frac{4m_\pi^4 }{3s(s-4m_\pi^2)}
    +\frac{m_{{\bar s}s}^2}{12(m_K^2-m_\pi^2)^2}\Bigl(m_\pi^2A_{10}-m_{{\bar s}s}^2A(m_{{\bar s}s})\Bigr)\r.\nn\\
& &\ -\frac{[8(m_K^2-m_\pi^2)^2(4m_\pi^6+22m_\pi^4
s-10m_\pi^2s^2+s^3)+
m_\pi^4s(s-4m_\pi^2)^2]}{12s(m_K^2-m_\pi^2)^2(s-4m_\pi^2)^2}A(m_\pi)\nn\\
& &\ \l.-\frac{m_\pi^2m_{{\bar
s}s}^2+s(4m_K^2-5m_\pi^2)}{6s(m_K^2-m_\pi^2)}\ln{\frac{m_\pi^2}{\mu^2}}
    -\frac{m_{{\bar s}s}^2(s-m_\pi^2)}{6s(m_K^2-m_\pi^2)}\ln{\frac{m_{{\bar s}s}^2}{\mu^2}}\r] \, ,\nn
\eea with \bea m_{\bar s s}^2&=&2m_K^2-m_\pi^2 \, , \nonumber \\
A_{10}&=&\Lambda \;
\ln{\frac{1-2\frac{m_K^2}{s}+\Lambda}{1-2\frac{m_K^2}{s}-\Lambda}}
- i \pi \theta\l(\sqrt{s}-(m_{{\bar s}s}+m_\pi)\r) \, , \nonumber
\\
\Lambda&=&\sqrt{1+4\frac{m_K^4}{s^2}+4\frac{m_\pi^4}{s^2}-4\frac{m_K^2}{s}(1+2\frac{m_\pi^2}{s})}\,
. \nonumber \eea The imaginary part of the amplitude diverges at
threshold, i.e. as $s \to 4 m_\pi^2$~\cite{BGLSU,Colangelo:1997ch}.

\section{Finite-Volume Scalar Correlation Function in Full QCD}
\label{app:fvsa} In this appendix we give the complete one-loop
expression of the finite-volume correlation function of the scalar
operator with two pion fields in the full theory:
\beq \bra{0}|
\pi^+_{- \vec q}(t_1) \pi^-_{\vec q}(t_2) S(0)\ket{0} =
\frac{e^{-Et_1}}{2E}\frac{e^{-Et_2}}{2E}
\l(-\frac{8}{f^2}\r)\l[1+\frac{1}{(4\pi f)^2}(I_{z}+I_a)+I_b(t_{1},t_{2})\r]\, ,
\label{eq:scft} \eeq where $E=\sqrt{\vec q^{2} +m^{2}_{\pi}}$.
Since we are interested in finite volume corrections we give only
the explicit expression for $I_b(t_1,t_2)$. We write \beq
I_b(t_{1},t_{2})= -\frac{E^2}{2 f^2}\l[A_{00}+A_{11}+A_{22}\r] \,
,\label{eq:ibu}\eeq where \bea A_{00}&=&\frac{m_\pi^2}{6E^2}
\frac{1}{L^3} \sum_{\vec
k}\frac{1}{w_0^2}\l\{\frac{1}{2(E-w_0)}-\frac{1}{2(E+w_0)} \r. \nn
\\
    &&\ \l.-e^{2(E-w_0)t_2}\l(\frac{1}{2(E-w_0)}+\frac{1}{2w_0}\r)+e^{2Et_2-2w_0t_1}
    \l(\frac{1}{2w_0}-\frac{1}{2(w_0+E)}\r)\r\} \, , \nn  \\
A_{11}&=&\frac{1}{L^3} \sum_{\vec k}\l\{\frac{d_{+}(w_1)}{2(E-w_1)}-\frac{d_{-}(w_1)}{2(E+w_1)}\r.  \nn \\
    &&\ \l.-e^{2(E-w_1)t_2}\l(\frac{d_{+}(w_1)}{2(E-w_1)}+\frac{d_{0}(w_1)}{2w_1}\r)
        +e^{2Et_2-2w_1t_1}\l(\frac{d_{0}(w_1)}{2w_1}-\frac{d_{-}(w_1)}{2(w_1+E)}\r)\r\} \nn \\
A_{22}&=&\frac{1}{L^3} \sum_{\vec k}\l\{\frac{c_{+}(w_2)}{2(E-w_2)}-\frac{c_{-}(w_2)}{2(E+w_2)}\r.  \nn \\
    &&\ \l.-e^{2(E-w_2)t_2}\l(\frac{c_{+}(w_2)}{2(E-w_2)}+\frac{c_{0}(w_2)}{2w_2}\r)
        +e^{2Et_2-2w_2t_1}\l(\frac{c_{0}(w_2)}{2w_2}-\frac{c_{-}(w_2)}{2(w_2+E)}\r)\r\}\,. \nn
\eea In the above expressions \bea w_{0}&=&\sqrt{\vec k^{2}
+m^{2}_{\bar s s}}\,; \quad w_{1}=\sqrt{\vec k^{2}
+m^{2}_{\pi}}\,; \quad w_{2}=\sqrt{\vec k^{2} +m^{2}_{K}}\,;
\nonumber \\
 c_{\pm}(w)&=&\frac{2}{3}\frac{E^2 \pm Ew+w^2}{E^2w^2}\,;\quad
c_{0}(w)=\frac{2}{3}\frac{1}{E^2}\,;
 \quad d_{\pm,0}(w)=2c_{\pm,0}(w)-\frac{m_\pi^2}{2E^2w^2}\,  . \nn
\eea Evaluating the sums, we find \beq \bra{0}|\pi^+_{-\vec
q}(t_1)\pi^-_{\vec q}(t_2)S(0)\ket{0}=
\frac{e^{-Et_1}}{2E}\frac{e^{-Et_2}}{2E}\l(-\frac{8}{f^2}\r)\l[1+{\rm
Re}({\cal A}_{\infty})+{\cal T}(t_{2})\r]\,, \eeq where, for
$E<m_K
< m_{\bar s s}$,  \bea {\cal T}(t_{2})&=&-\frac{E^2}{2f^2}
\l[-\frac{\nu t_2}{E^2
L^3}(4-\frac{m_\pi^2}{2E^2})+(2-\frac{3m_\pi^2}{4E^2})\l(-\frac{\nu}{3E^3L^3};-\frac{\nu}{6E^3L^3}\r)
\r.
  \nn \\ & & \l.  +\l(\frac{z(0)}{E^3L^3}(1-\frac{3m_\pi^2}{8E^2})-\frac{z(1)}{\pi^2EL}(1-\frac{m_\pi^2}{8E^2})\r)\r]
\label{eq:sopra} \eea and $\nu = \sum_{\vec k:w=E}$. We have used
\bea   z(s) =  \sum_{|\vec{l}| \neq
|\vec{n}|}\frac{1}{(\vec{l}^2-\vec{n}^2)^{s}}\quad \quad
\textrm{and}\quad\quad\vec q = \frac{2\pi}{L}\vec{n} \, .
\label{edefwow} \eea When we take the large-volume limit at fixed
two-pion energy, $W$, we expect that $z(s)$ scales as $L^{(2-2
s)}$ (and that $z(0)\sim -\nu \sim L^2 $). Thus the finite volume corrections
decrease as $1/L$ when $L \to \infty$~\cite{GVlatt02}.

The terms $\l(...;...\r)$ are the forward time contributions
($FTCs$) for the two cases $t_1 \gg t_2$ and $t_2=t_1$
respectively. These are the terms which are reabsorbed by the sink
when the matrix element is extracted (for a detailed discussion
see section 4.1 of ref.~\cite{spqrdi32logs}).

From the above equations we find \bea \Delta
W&=&-\frac{\nu}{4f^2L^3}(8-\frac{m_\pi^2}{E^2})\rightarrow-\frac{7}{4f^2L^3}
\Rightarrow a_{0}=\frac{7m_\pi}{16\pi f^2} \nn \\
\sqrt{LL}&=&1-\frac{E^2}{2f^2}\l[\frac{\nu}{(EL)^3}\l(\frac{3m_\pi^2}{8E^2}-1\r)
-\frac{z(1)}{\pi^2EL}\l(1-\frac{m_\pi^2}{8E^2}\r)\r]\rightarrow
    1+\frac{m_\pi^2}{16\pi^2f^2}\l(\frac{7z(1)}{m_\pi L}+\frac{5\pi^2}{(m_\pi L)^3}\r) \nn \\
FTCs&=&1+\l(1-\frac{3m_\pi^2}{8E^2}\r)\l(\frac{\nu}{3f^2EL^3};\frac{\nu}{6f^2EL^3}\r)
\rightarrow 1+\l(\frac{5}{24f^2m_\pi L^3};\frac{5}{48f^2m_\pi
L^3}\r) \, , \label{eq:varie}\eea where the limits refer to the
case with the two pions at rest. All these results are in
agreement with expectations: the energy shift in a finite volume
is precisely the one predicted by the L\"uscher quantization
condition~\cite{luscher}; the $LL$ factor is in agreement with the
general formula of ref.~\cite{LL} and the $FTCs$ term will be
cancelled when we divide the correlation function by the square
root of the $\pi\pi \to \pi\pi$ correlator of eq.~(\ref{eq:4pi}).

\section{$\!$Finite-Volume Scalar Correlation Function in Quenched QCD}
\label{app:fvqsa} In this appendix we give the complete one-loop
expression of the finite-volume correlation function for the
scalar operator with two pion fields in the quenched theory: \beq
\bra{0}| \pi^+_{- \vec q}(t_1)\pi^-_{\vec q}(t_2)S^q(0) \ket{0} =
\frac{e^{-Et_1}}{2E}\frac{e^{-Et_2}}{2E} \l(-\frac{8}{f^2}\r)\l[1+
 \frac{1}{(4\pi f)^2}(I^q_z + I^q_a)+I^q_b(t_1,t_2)\r]\, , \\ \label{eq:scqt}
\eeq where
\begin{equation}
I^q_b(t_1,t_2)=-\frac{E^2}{2 f^2}\l[A_{00}+A_{10}+A_{11}+A_{22}\r]
\, ,\eeq with \bea A_{00}&=&a_{00} \frac{1}{L^3} \sum_{\vec
k}\frac{1}{w_0^2}\l\{\frac{1}{2(E-w_0)}-\frac{1}{2(E+w_0)} \r. \nn
\\
    &&\ \l.-e^{2(E-w_0)t_2}\l(\frac{1}{2(E-w_0)}+\frac{1}{2w_0}\r)+e^{2Et_2-2w_0t_1}
    \l(\frac{1}{2w_0}-\frac{1}{2(w_0+E)}\r)\r\} \, , \nn  \\
      &&+\frac{1}{L^3} \sum_{\vec k}\frac{1}{w_0^2}\l\{\frac{b}{2(E-w_0)}-\frac{b}{2(E+w_0)}\r. \nn \\
    &&\ \l.-e^{2(E-w_0)t_2}\l(\frac{b}{2(E-w_0)}+\frac{\tilde b}{2w_0}\r)+e^{2Et_2-2w_0t_1}\l(\frac{\tilde b}{2w_0}-\frac{b}{2(w_0+E)}\r)\r\} \, , \nn  \\
A_{10}&=&a_{10} \frac{1}{L^3} \sum_{\vec k}\frac{1}{w_0w_1}\l\{\frac{1}{2E-w_0-w_1}-\frac{1}{2E+w_0+w_1}\r. \nn \\
    &&\
    \l.-e^{(2E-w_0-w_1)t_2}\l(\frac{1}{2E-w_0-w_1}+\frac{1}{w_0+w_1}\r)\r.\nn\\
    &&\l.\
        +e^{2Et_2-(w_0+w_1)t_1}\l(\frac{1}{w_0+w_1}-\frac{1}{w_0+w_1+2E}\r)\r\}\,,\nn\\
A_{11}&=&\frac{1}{L^3} \sum_{\vec k}\l\{\frac{c_{+}(w_1)}{2(E-w_1)}-\frac{c_{-}(w_1)}{2(E+w_1)}\r.  \nn \\
    &&\ \l.-e^{2(E-w_1)t_2}\l(\frac{c_{+}(w_1)}{2(E-w_1)}+\frac{c_{0}(w_1)}{2w_1}\r)
        +e^{2Et_2-2w_1t_1}\l(\frac{c_{0}(w_1)}{2w_1}-\frac{c_{-}(w_1)}{2(w_1+E)}\r)\r\} \nn \\
    &&+\frac{2}{L^3} \sum_{\vec k}\frac{1}{w_1^2}\l\{\frac{b}{2(E-w_1)}-\frac{b}{2(E+w_1)}\r.  \nn \\
    &&\ \l.-e^{2(E-w_1)t_2}\l(\frac{b}{2(E-w_1)}+\frac{\tilde b}{2w_1}\r)+e^{2Et_2-2w_1t_1}\l(\frac{\tilde b}{2w_1}-\frac{b}{2(w_1+E)}\r)\r\} \nn  \\
    &&+\frac{1}{L^3} \sum_{\vec k}\frac{a_{11}^{(1)}(w)}{E^2}\l\{\frac{1}{2(E-w_1)}-\frac{1}{2(E+w_1)}\r.  \nn \\
    &&\ \l.-e^{2(E-w_1)t_2}\l(\frac{1}{2(E-w_1)}+\frac{1}{2w_1}\r)+e^{2Et_2-2w_1t_1}\l(\frac{1}{2w_1}-\frac{1}{2(w_1+E)}\r)\r\} \nn  \\
    &&+\frac{1}{L^3} \sum_{\vec k}\frac{a_{11}^{(2)}(w)}{E}\l\{\frac{1}{[2(E-w_1)]^2}+\frac{1}{[2(E+w_1)]^2}\r.  \nn \\
    &&\ \l.-e^{2(E-w_1)t_2}\l(\frac{1-[2(E-w_1)]
    t_2}{[2(E-w_1)]^2}-\frac{1+[2w_1]t_2}{[2w_1]^2}\r)\r.\nn\\ &&\
    \l.
        -e^{2Et_2-2w_1t_1}\l(\frac{1+[2w_1]t_1}{[2w_1]^2}-\frac{1+[2(w_1+E)]t_1}{[2(w_1+E)]^2}\r)\r\} \nn  \\
    &&+\frac{1}{L^3} \sum_{\vec k}a_{11}^{(3)}(w)\l\{\frac{1}{[2(E-w_1)]^3}-\frac{1}{[2(E+w_1)]^3} \r. \nn \\
    &&\ \l.-e^{2(E-w_1)t_2}\l(\frac{1-[2(E-w_1)] t_2+[2(E-w_1)]^2 t_2^2/2}{[2(E-w_1)]^3}
        +\frac{1+[2w_1]t_2+[2w_1]^2 t_2^2/2}{[2w_1]^3}\r)\r. \nn \\
    &&\ \l.+e^{2Et_2-2w_1t_1}\l(\frac{1+[2w_1]t_1+[2w_1]^2t_1^2/2}{[2w_1]^3}
        -\frac{1+[2(w_1+E)]t_1+[2(w_1+E)]^2t_1^2/2}{[2(w_1+E)]^3}\r)\r\}\,,\nn\\
A_{22}&=&\frac{1}{2L^3} \sum_{\vec k}\l\{\frac{c_{+}(w_2)}{2(E-w_2)}-\frac{c_{-}(w_2)}{2(E+w_2)}\r.  \nn \\
    &&\ \l.-e^{2(E-w_2)t_2}\l(\frac{c_{+}(w_2)}{2(E-w_2)}+\frac{c_{0}(w_2)}{2w_2}\r)
        +e^{2Et_2-2w_2t_1}\l(\frac{c_{0}(w_2)}{2w_2}-\frac{c_{-}(w_2)}{2(w_2+E)}\r)\r\}\,. \nn
\eea In the above equations
 \bea b&=&\tilde b -\frac23 v_1 \,,\quad \tilde
b=\frac{m_\pi^2(v_1-v_2)}{3E^2}\,, \nn \\
a_{00}&=&\frac{y[m_{0}^2-\alpha m_K^2(2-y)]^2}{36 E^2 m_K^2
(1-y)^2} \, , \nn \\ a_{10}&=&-\frac{y[m_{0}^2-\alpha m_K^2
y][m_{0}^2-\alpha m_K^2 (2-y)]}{18 E^2 m_K^2 (1-y)^2} \, , \nn \\
a_{11}^{(1)}(w_1)&=&-\frac{m_{0}^2 m_K^2 y}{3 w_1^4}+\frac{\alpha
m_K^2 y(y m_K^2-2w_1^2)}{3 w_1^4} \nn \\
        & &+\frac{m_{0}^4 y [w_1^4+2 m_K^4(1-y)^2]-2 \alpha m_{0}^2 y m_K^2[(1-y)^2(-4m_K^2w_1^2+2m_K^4 y)+y w_1^4]}{36 m_K^2 w_1^6(1-y)^2} \nn\\
        & &+\frac{\alpha^2 m_K^4 y[2m_K^2 y(1-y)^2(-4w_1^2+ym_K^2)+w_1^4(8-16y+9y^2)]}{36 m_K^2 w_1^6(1-y)^2}\, , \nn \\
a_{11}^{(2)}(w_1)&=&-\frac{y m_K^2 [m_{0}^2-\alpha m_K^2
y][m_{0}^2-\alpha m_K^2 y +(2 \alpha-3)w_1^2]}{9 E w_1^5}\, , \nn
\\ a_{11}^{(3)}(w_1)&=&\frac{y m_K^2 [m_{0}^2-\alpha m_K^2 y]^2}{9
E^2 w_1^4}\, . \nn \eea where $y=m_\pi^2/m_K^2$ and $m_{0}$ and
$\alpha$ are the parameters characterizing the $\eta^\prime$
propagator~\cite{spqrdi32logs}. Evaluating the sums, we obtain:
\beq \bra{0}|\pi^+_{-\vec q}(t_1)\pi^-_{\vec
q}(t_2)S^q(0)\ket{0}=\frac{e^{-Et_1}}{2E}\frac{e^{-Et_2}}{2E}\l(-\frac{8}{f^2}\r)\l[1+{\rm
Re}({\cal A}^q_{\infty})+ {\cal T}^q(t_2)\r]\,,\nn \\ \eeq where,
for $E \le m_K \le m_{\bar s s }$ we have \bea {\cal
T}^q(t_2)&=&-\frac{E^2}{2f^2}\l\{ \l[-\frac{2\nu t_2}{E^2
L^3}+\l(-\frac{\nu}{3E^3L^3};-\frac{\nu}{6E^3L^3}\r)
    +\l(\frac{z(0)}{2E^3L^3}-\frac{z(1)}{2\pi^2EL}\r)\r] \r. \nn \\
& &\l.+\l[-\frac{2 b \nu t_2}{E^2 L^3}+\l(-\frac{\nu \tilde b}{E^3L^3};-\frac{\nu b}{2E^3L^3}\r)
    +\l(\frac{3 b z(0)}{2E^3L^3}-\frac{b z(1)}{2\pi^2EL}\r)\r] \r. \nn \\
& &\l.+ a_{11}^{(1)}(E)\l[-\frac{\nu t_2}{E^2 L^3}+\l(-\frac{\nu}{2E^3L^3};-\frac{\nu}{4E^3L^3}\r)
    +\l(\frac{b_{11}^{(1)}(E)}{a_{11}^{(1)}(E)}\frac{z(0)}{4E^3L^3}-\frac{z(1)}{4\pi^2EL}\r)\r] \r.+ \nn \\
& &\l.\hspace{-20pt}+ a_{11}^{(2)}(E)\l[\frac{\nu t_2^2/2}{E
L^3}+\l(\frac{(1+2Et_2)\nu}{4E^3L^3};\frac{(1+4Et)\nu}{16E^3L^3}\r)
    +\l(\frac{c_{11}^{(2)}(E)z(0)}{16E^3L^3}+\frac{b_{11}^{(2)}(E)z(1)}{4\pi^2EL}+\frac{z(2)}{16\pi^4}EL\r)\r] \r. \nn \label{eq:calTq} \\
& &\l.+ a_{11}^{(3)}(E)\l[\frac{\nu t_2^3/6}{L^3}+\l(-\frac{(1+2Et_2+2E^2t_2^2)\nu}{8E^3L^3};-\frac{(1+4Et+8E^2t^2)\nu}{64E^3L^3}\r)
\r. \r. \nn \\
& & \l.
+ \l(\frac{111 z(0)}{64E^3L^3}-\frac{3 z(1)}{8\pi^2 E L}+\frac{5 z(2)}{64\pi^4}EL-\frac{z(3)}{64\pi^6}(EL)^3\r)\r]
\, ; \\
b_{11}^{(1)}(E)&=&-\frac{7m_{0}^2 m_K^2 y}{3 E^4}+\frac{\alpha m_K^2 y(7y m_K^2-6E^2)}{3 E^4} \nn \\
        & &+\frac{m_{0}^4 y [3E^4+22 m_K^4(1-y)^2]-2 \alpha m_{0}^2 y m_K^2[(1-y)^2(-28m_K^2E^2+22m_K^4 y)+3y E^4]}{36 m_K^2 E^6(1-y)^2} \nn\\
        & &+\frac{\alpha^2 m_K^4 y[2m_K^2 y(1-y)^2(-28E^2+11ym_K^2)+3E^4(8-16y+9y^2)]}{36 m_K^2 E^6(1-y)^2}\, , \nn \\
c_{11}^{(2)}(E)&=&\frac{17 E^2(2\alpha-3)+49m_0^2-49y\alpha m_K^2}{E^2(2\alpha-3)+m_0^2-y\alpha m_K^2} \, , \nn \\
b_{11}^{(2)}(E)&=&-\frac{E^2(2\alpha -3)+2m_0^2-2y\alpha m_K^2}{E^2(2\alpha-3)+m_0^2-y\alpha m_K^2} \, . \nn
\eea


\end{document}